# The use of the Rietveld method and pairs distribution function analysis to study the pressure dependence of the trigonal $SnSe_2$ and $SnS_2$ structure

J. C. de Lima[1,a)], Z. V. Borges[1], C. M. Poffo[2], S. M. Souza[3], D. M. Trichês[3], and R.S. de Biasi[4]

[1]*Departamento de Engenharia Mecânica, Universidade Federal de Santa Catarina, Programa de Pós-Graduação em Ciência e Engenharia de Materiais, Campus Universitário Trindade, C.P. 476, 88040-900 Florianópolis, Santa Catarina, Brazil*

[2]*Universidade Federal de Santa Catarina, Campus de Araranguá, 88900-000 Araranguá, Santa Catarina, Brazil.*

[3]*Departamento de Física, Instituto de Ciências Exatas, Universidade Federal do Amazonas, 69077-000 Manaus, Amazonas, Brazil*

[4]*Seção de Engenharia Mecânica e de Materiais, Instituto Militar de Engenharia, 22290-270 Rio de Janeiro, RJ, Brazil*

ABSTRACT

A nanostructured trigonal $SnSe_2$ was produced by mechanical alloying, and the effect of high-pressures up to 25.8 GPa on it was investigated. The literature reports refined structural data for $SnS_2$ for pressures up to 20 GPa. These data were used as input data in a crystallographic software to calculate the shell structures around the Sn, Se, and S atoms placed at the origin. The shell structures were used to simulate the partial and total structure factors $S_{i-j}(K)$ and $S(K)$, and by Fourier transformation the partial and total pairs distribution functions $G_{i-j}(R)$ and $G(R)$ were obtained. The effect of high-pressure on the $SnSe_2$ and $SnS_2$ structures were followed by observing the changes in the $G_{i-j}(R)$ functions. Also, the effect of high-pressure on the smallest angle Sn-X (X=Se,S)-Sn, intralayer distance Sn-X (X=Se,S), and interlayers distance X-X (X=Se,S) was studied. The interlayers distance X-X (X=Se,S) changes faster than the intralayer distance Sn-X (X=Se,S). An enhancement of the average power factor at 20 GPa and 800 K for $SnS_2$ was reported. Using the interlayers distance S-S and intralayer distance Sn-S, it was evidenced that the enhancement of average power factor may be associated with the changes of the interlayers distance S-S that is faster than that of the intralayer distance Sn-S.

*Keywords*: High pressure; X-ray diffraction; Rietveld Method; Pairs Distribution Function; Thermoelectric layered compoud

a) Author to whom correspondence should be addressed. Electronic mail: joao.cardoso.lima@ufsc.br or jcardoso.delima@gmail.com

## I. Introduction

Despite the extraordinary progress in the development of DACs, allowing that several dozen GPa be achieved, they have yet a limited window of few tens (two or three) of degrees, imposing a restriction in the angular range ($2\theta$) in which the X-ray diffraction (XRD) patterns are measured. This limitation makes difficult to visualize the effect of high-pressure on the structure being studied, as well as it increases the difficulties in the determination of new high-pressure structure phases emerging due to the few number of diffraction peaks recorded.

Trying to overcome the difficulties mentioned above, at least partially, we implemented an approach combining the Rietveld method (RM) [1] and pairs distribution function (PDF) analysis [2−4]. The details of this approach are described in Refs. [5,6,7], and it will not be repeated here. RM has been implemented in several computational packages, such as the GSAS package [8], which has been employed to refine and/or to determine structures of the crystallographic phases of single crystal and polycrystalline solids. The theoretical description of PDF analysis as well as its applications to the liquid and amorphous materials is well documented in the literature [2−4] and will not be repeated here. The approach to be presented in this study requires only the knowledge of the structure of the crystalline material. It has already been used by us to investigate the effect of moderate (MPa) and high (GPa) pressures on nanocrystalline materials [6,7]. In order to corroborate the validity of this approach and also make it more explicit for readers, it was also applied to the $Ta_2O_5$ orthorhombic compound previously studied using XRD, RM, and PDF [9,10]. The results obtained using this approach were compared with those reported in Refs. [9,10], and an excellent agreement was observed.

The isostructural $SnSe_2$ and $SnS_2$ compounds (space group P-3 m 1, 164) and based on $SnS_2$-, $SnSe_2$-alloys have great potential for applications in thermoelectric (TE) devices like solid-state coolers or generators at moderate and high temperatures [9]. Recently, our group produced using the mechanical alloying (MA) technique the $SnSe_2$ compound, in the nanostructured form.

The effect of high-pressure up to 25.8 GPa on $SnSe_2$ was investigated using X-ray diffraction (XRD) measurements and synchrotron radiation. The recorded XRD patterns were refined using RM. The refined structural data are used as "input data" in a crystallographic software such as Crystal Office 98® from the Atomic Softek - Canada, and using the option (tool "output + coordinates + shell structure") the shell structure around a specific atom $i$ placed at the origin is obtained. Recently, Filsø *et al*. [11], using single crystal

XRD measurements and DFT calculations, investigated the effect of high-pressure up to 20 GPa on the $SnS_2$ compound. The recorded XRD patterns were refined using RM.

Javed *et al*. [12], using first principles calculations, studied along *a*- and *c*-axes, the thermoelectric power (S), conductivity electrical (σ), and power factor (PF) properties for atmospheric, 10 and 20 GPa, at temperatures 300 K, 500 K and 800 K. Those researchers reported an enhancement in average PF at 20 GPa and 800 K, leading to values of figure of merit ZT between 1 − 2.5. TE materials are characterized for technological applications by the dimensionless figure of merit ZT. To our best knowledge, similar study was not found in the literature for $SnSe_2$.

As mentioned previously, the approach combining the Rietveld method and pairs distribution function analysis requires only the knowledge of the structure of the crystalline material to build the partial pairs distribution functions (PPDFs) as well as the total pairs distribution function (PDF). The PDF functions can give a more accurate insight structure of the effect of high-pressure. We are using the refined structural data for $SnSe_2$ to calculate the shell structures around Sn and Se atoms placed at the origin and those reported for $SnS_2$ to calculate the shell structures around Sn and S atoms. Thus, the main goal of this study is to compare the PPDFs and PDFs for the $SnS_2$ and $SnSe_2$ compounds. In addition, it will be tried to associate, at least partially, the structural changes promoted by high-pressure on $SnS_2$ with the changes observed in the macroscopic physical thermoelectric power (S), conductivity electrical (σ), and power factor (PF) properties.

**II. Experimental procedure**

The nanostructured $SnSe_2$ powder used in this study was the same used in a previous study [13]. Then, details involving its production are described in [13] and will not be repeated here. *In-situ* XRD patterns were recorded for pressures listed in Table 1, at atmospheric pressure, at the X-ray Diffraction and Spectroscopy (XDS) station of the LNLS synchrotron radiation facility (Brazil). The XDS station consists of a nitrogen-cooled Si (311) double-crystal monochromator to select the incoming energy and a Dectris Pilatus 300 K image plate detector to record the diffracted beam. The present experiment was performed using a wavelength of λ = 0.61990 Å (20.0 keV). The sample was loaded in a stainless-steel gasket hole of 150 μm diameter. The diamond culet was 350 $\mu$m in diameter and the thickness of the gasket after indentation was 65 $\mu$m. As the pressure-transmitting medium, a methanol-ethanol (4:1) solution was used, which is hydrostatic in the 0−10 GPa pressure range and nonhydrostatic for larger pressures. A Membrane Diamond Anvil Cell (MDAC) with an opening of 22.0° (2$\theta$) was used. The sample-to-detector distance (292 mm) and instrumental broadening were determined from diffraction data of standard crystalline $LaB_6$ powder. The pressure was determined by the fluorescence shift of a ruby micropiece [14] loaded in the sample chamber. An exposure time of 10 min was used for all measurements. The two-dimensional diffraction patterns were converted to intensity versus 2$\theta$ data using the FIT2D software [15]. All the XRD patterns were refined using the Rietveld method [1] implemented in the computational GSAS package [8]. For $SnS_2$, all the experimental details about the preparation of sample, XRD measurements, pressure applied, and data refinement are described in Ref. [11], and they will not be repeated here.

**III. The use of the structural data for $SnSe_2$ and $SnS_2$ to build the shell structure**

At room temperature and atmospheric pressure, the isostructural $SnS_2$ and $SnSe_2$ compounds crystallize in a trigonal structure (space group P-3 m 1, No. 164), with the Sn atoms occupying the Wyckoff position *1a* (0,0,0) and S, Se atoms occupying the position *2d* (1/3, 2/3, Z) [16]. Figure 1 shows the conventional unit cell for Sn(S,Se)$_2$. These compounds are characterized by intralayer covalent Sn–Se, Sn–S bonds and interlayer Se···Se, S···S of the van der Waals type. As previously mentioned, the effect of high-pressure on SnSe$_2$ for pressures up to 25.8 GPa was studied using XRD measurements, and the recorded patterns were refined using RM [1] implemented in the GSAS package [8]. Table 1 lists the refined lattice parameters and the Z-coordinate for the Se atoms at position *2d* (1/3, 2/3, Z). For SnS$_2$, these structural data were taken from the paper by Filsø *et al.*[11], and they are listed in Table 2.

The space group P-3 m 1 (164), lattice parameters for SnSe$_2$ (Table 1) and for SnS$_2$ (Table 2), Wickoff positions *1a* (0,0,0) for the Sn atoms and *2d* (1/3,2/3,Z) for the Se, S atoms were used as "input data" for the Crystal Office software® [17], and using the software´s tools "*Output + Coordinates + Shell Structure*" the shell structures around Sn and Se atoms placed at the origin for SnSe$_2$, and around Sn and S atoms placed at the origin for SnS$_2$ were calculated up to *Rmax* = 25 Å (arbitrarily chosen). The shell structures give the coordination numbers $N_{i-j}$ and interatomic distances $R_{i-j}$ with respect to the atom *i* placed at the origin, i.e., $N_{Sn-Sn}$, $R_{Sn-Sn}$, $N_{Sn-Se}$, $R_{Sn-Se}$, $N_{Se-Se}$, $R_{Se-Se}$ for SnSe$_2$, and $N_{Sn-Sn}$, $R_{Sn-Sn}$, $N_{Sn-S}$, $R_{Sn-S}$, and $N_{S-S}$, $R_{S-S}$ for SnS$_2$. The shell structures were used as "input data" for a computational FORTRAN code describing the expression (1) of Ref. [5] to simulate the partial structure factors $S_{i-j}(K)$, i.e., $S_{Sn-Sn}(K)$, $S_{Sn-Se}(K)$, $S_{Se-Se}(K)$ for SnSe$_2$, and $S_{Sn-Sn}(K)$, $S_{Sn-S}(K)$, $S_{S-S}(K)$ for SnS$_2$, using a value *Kmax* = 30 Å$^{-1}$ arbitrarily chosen. By Fourier transformation of the simulated $S_{i-j}(K)$ factors, the partial and total reduced total distribution functions $\gamma_{i-j}(R)$ and $\gamma(R)$, the partial and total pairs distribution functions $G_{i-j}(R)$ and $G(R)$, and partial and total radial distribution functions $RDF_{i-j}(R)$ and $RDF(R)$ were obtained. To build the total $S_{SnSe2}(K)$ and $S_{SnS2}(K)$ factors, the weights $W_{Sn-Sn}(K)$, $W_{Sn-Se}(K)$, $W_{Se-Se}(K)$, $W_{Sn-S}(K)$ and $W_{S-S}(K)$ were computed. All the computational FORTRAN codes describing the expressions given in Section II of Ref. [5] were written by one of the authors (J.C. de Lima).

**IV. Results and discussion**

Figures 2(a)-(b) show the $\gamma_{SnSe2}(R)$ functions obtained by Fourier transformation of the simulated $S_{SnSe2}(K)$ factors, at ambient conditions (T = 300 K, P = 0 GPa), with different values of $Kmax$. In this figure is also shown the experimental $\gamma_{SnSe2}(R)$ function (green curve in Fig. 2(a)) obtained by Fourier transformation of the experimental $S_{SnSe2}(K)$ factor derived from the XRD data recorded at ambient conditions with the SnSe$_2$ powder within the DACs. In order to reproduce the angular range reached using a DAC, a conventional source (Cu and Mo target), and a synchrotron source, the values of $Kmax$ 5, 8, 16, 30 and 60 Å$^{-1}$, with a step $\Delta K$= 0.025 Å$^{-1}$, were considered to obtain the $S_{i-j}(K)$ factors. For simulations, the lattice parameters for pressure P = 0 GPa listed in Table 1 were considered. Based on Ref. [5], the value of $Kmax$ 30 Å$^{-1}$ was assumed in this study. The resolution of $\gamma(R)$ function is given by $\Delta R = 3.8/Kmax$ [3], and for $Kmax$ 30 Å$^{-1}$, $\Delta R = 0.127$ Å. From these figures one can see that the approach combining RM and PDF analysis can give a insight structure of the effect of high-pressure, as shown in Fig. 2(b). However, accurate high-pressure XRD measurements and accurate determination of the structure are necessary.

Figures 3(a)-(c), (d)-(f), and (g)-(i) show the $G_{Sn-Sn}(R)$, $G_{Sn-Se}(R)+G_{Sn-S}(R)$, and $G_{Se-Se}(R)+G_{S-S}(R)$ functions for SnSe$_2$ and SnS$_2$, respectively, obtained by Fourier transformation of their respective simulated $S_{i-j}(K)$ with the value of $Kmax$ 30 Å$^{-1}$. According to the calculated shell structures for SnSe$_2$, the first coordination shells of the $G_{Sn-Sn}(R)$, $G_{Sn-Se}(R)$, and $G_{Se-}$

$_{Se}(R)$ functions are formed by 6 Sn-Sn pairs at 3.811 Å, 6 Sn-Se pairs at 2.682 Å, and two subshells each one containing 6 Se-Se pairs at 3.776 Å and 3.881 Å, respectively. For $SnS_2$, at ambient conditions, the first coordination shells of the $G_{Sn-Sn}(R)$, $G_{Sn-S}(R)$, and $G_{S-S}(R)$ functions are formed by 6 Sn-Sn pairs at 3.646 Å, 6 Sn-S pairs at 2.560 Å, and two subshells containing 3 S-S pairs at 3.595 Å and 9 S-S pairs at 3.646 Å, respectively. With increasing pressure, the effect of pressure on $G_{Sn-Sn}(R)$ of $SnSe_2$ and $SnS_2$ is only visible for the coordination shell located between ≈ 6.3 Å < R < ≈ 7.3 Å, where the two subshells tend to overlap; On $G_{Sn-Se}(R)$ and $G_{Sn-S}(R)$ of $SnSe_2$ and $SnS_2$, with exception of the first coordination shell, the effect is to overlap the subshells forming the coordination shells. This effect seems to be more significant for $SnSe_2$; and on $G_{Se-Se}(R)$ and $G_{S-S}(R)$ of $SnSe_2$ and $SnS_2$, the effect is to split the two subshells forming the first coordination shell. This effect seems to be more significant for $SnS_2$. Simulated total $G_{SnSe2}(R)$ and $G_{SnS2}(R)$ for the pressures used are shown in Fig. 4(a)-(c). With increasing pressure, no splitting is observed for the first coordination shell of $SnSe_2$ and $SnS_2$; The second coordination shell of $SnSe_2$ is splitted into two subshells partially overlapped, while that of $SnS_2$, no splitting is observed.

Figure 5 shows the pressure dependence of the smallest angle Sn-X(X=Se,S)-Sn, intralayer interatomic distances Sn-X(X=Se,S), and interlayers interatomic distances X-X (X=Se,S). From this figure, one can see that, with increasing pressure, the value of the smallest angle Sn-Se-Sn increase for the pressures up to ≈18 GPa, decreasing slightly for larger pressures, while the smallest angle Sn-S-Sn decreases for the pressures used. With respect to the intralayer distances Sn-Se and Sn-S, with increasing pressure, they decrease for the pressures used. However, the later decreases faster. For the distance Sn-S, the increase between 12 and 14 GPa can be associated to the nonhydrostaticity of the pressure transmitting médium. With respect to the interlayers distance X-X (X=Se,S), with increasing pressure, they decrease continuously. For SnS$_2$, it decreases faster.

As previously mentioned, Javed *et al*. [12], using a first principles calculations, studied for SnS$_2$ the macroscopic thermoelectric power (S), conductivity electrical (σ), power factor (PF) properties along the *a*- and *c*-axes, and average PF at atmospheric pressure (P = 0 GPa), 10 GPa and 20 GPa and at temperatures 300 K, 500 K and 800 K. Their results are shown in Fig. 6(a)-(d). Thus, it is interesting to try to correlate the effect of high-pressure previously shown with the macroscopic thermoelectric properties of SnS$_2$. The performance of TE material is characterized by its dimensionless figure of merit ZT= (S$^2$σ)T/k, where S is the thermoelectric power, σ the electrical conductivity, k the termal conductivity, and T the absolute temperature. The S$^2$σ is named the power fator (PF). For technological TE applications, ZT > 1 is desired. Those researchers reported an enhancement of TE performance of SnS$_2$ at 20 GPa and 800 K (1 < ZT < 2.5). From the Fig. 6(a), one can see that, with increasing pressure, carrier concentration and temperature, the thermoelectric power S decreases. At atmospheric pressure and 800 K, the value of S along *a*-axis is larger than the value along *c*-axis. At 10 GPa and 800 K, the value of S along *a*-axis is slightly larger than the value along *c*-axis, and at 20 GPa and 800 K, the values of S alog *a*- and *c*-axes are similar. From the Fig. 6(b), one can see that, with increasing pressure, carrier concentration and temperature, the electrical conductivity σ increases. At atmospheric pressure and 800 K, the value of σ along *a*-axis is larger than the value along *c*-axis. At 10 GPa and 800 K, the values of σ along *a*- and *c*-axes are similar, and at 20 GPa and 800 K, the value of σ along *c*-axis is larger than the value of σ along *a*-axis. This behavior at 20 GPa and 800 K may be associated with the interlayers distance S-S that decreases faster than the Sn-S one, suggesting that the electronic repulsion begins to play an important role. From the Fig. 6(c), one can see that the power fator (PF) increases with increasing the carrier concentration and temperature, but it can decreases with increasing pressure. At atmospheric pressure and 800 K, the of PF along *a*-axis is larger than the value of PF along *c*-axis. This behavior seems to be in accordance with the values of S and σ along this axis. At 10 GPa and 800 K, value of PF along the *a*-axis is even larger than the value along *c*-axis, but the difference between them decreased, and at 20 GPa and 800 K, value of PF along the *c*-axis is larger than the value along *a*-axis. This behavior seems to be in accordance with the values of S and σ along this axis. Thus, the increase in PF along the *c*-axis may be associated with the interlayers distance S-S that decreases faster than the Sn-S one, suggesting that the electronic repulsion begins to play an important role, mainly, in the electrical conductivity. From the Fig. 6(d), one can see that the average power fator (PF) increases with increasing pressure and carrier concentration. At 20 GPa and 800 K, the increase in average PF may be associated to the increase in value of the electrical conductivity along *c*-axis due to the fast decrease in the interlayers distance S-S, as discussed early. To our best knowledge, there are no similar data for $SnSe_2$ in the literature, and therefore, similar analysis is inaccurate. From Fig. 5, one can see that, with increasing pressure, both the intrayer distance Sn-Se ($d_{intra}$) and interlayers interatomic distance Se-Se ($d_{inter}$) decrease continuously for $SnSe_2$, while for $SnS_2$ the intrayer distance Sn-S decrease slowly, whereas the interlayers interatomic distance S-S decreases fast and continuously. Individual distance compressibility rates $d_i = -(1/d_0)(\partial d_i/\partial P)_T$ ($i = d_{intralayer}, d_{interlayers}$) were calculated for both compounds. For $SnSe_2$, $d_{intralayer} = 0.0504$ GPa$^{-1}$ and $d_{interlayers} = 0.0853$ GPa$^{-1}$, while for $SnS_2$, $d_{intralayer} = 0.0232$ GPa$^{-1}$ and $d_{interlayers} = 0.1524$ GPa$^{-1}$. From these values, one can see that, for $SnSe_2$, the change in the interlayers distance Se-Se is 41 % faster than that in the intralayer distance Sn-Se, while for SnS$_2$, the interlayers distance S-S changes 85 % faster than the intralayer distance Sn-S. It is evidenced above that the increase in the average PF for SnS$_2$ at 20 GPa and at 800 K may be associated with the decrease of the interlayers distance S-S, increasing the electrical conductivity along the *c*-axis. If we assume that this is true, for SnSe$_2$ at the same conditions, a reduction of half in the average PF is expected. Therefore, the figure of merit ZT for SnSe$_2$ can be smaller than that for SnS$_2$ at 20 GPa and 800 K.

## V. CONCLUSIONS

This study reports the results obtained of application of an approach combining the Rietveld method and pairs distribution function analysis to study crystalline materials under high-pressure and/or temperature for trigonal SnSe$_2$ and SnS$_2$. The refined structural parameters obtained from the Rietveld refinement of the XRD patterns measured for SnSe$_2$ and for SnS$_2$ (taken from Ref. [11] were used as input data to simulate the partial and total structure factors $S_{Sn-Sn}(K)$, $S_{Sn-Se}(K)$, $S_{Sn-S}(K)$, $S_{Se-Se}(K)$, $S_{S-S}(K)$, $S_{SnSe2}(K)$, and $S_{SnS2}(K)$. Fourier transformation of them permitted to obtain the partial and total pairs distribution functions $G_{Sn-Sn}(R)$, $G_{Sn-Se}(R)$, $G_{Sn-S}(R)$, $G_{Se-Se}(R)$, $G_{S-S}(R)$, $G_{SnSe2}(R)$, and $G_{SnS2}(R)$. Using the structural data obtained from the Rietveld refinement, the pressure dependence of the smallest angle Sn-X (X=Se,S)-Sn, intralayer distance Sn-X (X=Se,S), and interlayers distance X-X (X=Se,S) were studied. For both the trigonal SnSe$_2$ and SnS$_2$ structures is shown that the interlayers distances change faster than the intralayer distances. Using calculated data for the thermoelectric power (S), electrical conductivity (σ), thermoelectric power factor (PF) along *a*- and *c*-axes, and average PF for SnS$_2$, and the interlayers and intralayer distances obtained from the refined structural data for SnS$_2$, it was evidenced that the enhancement of average power factor observed at 20 GPa and 800 K may be associated with the fast decrease of the interlayers distance X-X (X=Se,S).

ACKNOWLEDGMENTS

One of the authors (Z. V. Borges) was financially supported by a scholarship from CNPq.

TABLES

TABLE 1: Lattice parameters, volume, z atomic coordinate of site 2d, reduced $\chi^2$ and R factors obtained from the Rietveld refinements for the SnSe$_2$ trigonal phase for the applied pressures.

| P (GPa) | a (Å) | c (Å) | V (Å$^3$) | z (Se) | $R_{wp}$ (%) | $R_p$ (%) | $\chi^2$ |
|---|---|---|---|---|---|---|---|
| 0.0(1) | 3.811(1) | 6.13(7) | 77.19(1) | 0.25 | 5.81 | 4.7 | 6.895 |
| 2.8(1) | 3.757(2) | 5.86(4) | 71.69(3) | 0.250(7) | 3.44 | 2.43 | 0.1086 |
| 3.6(1) | 3.745(1) | 5.81(3) | 70.61(6) | 0.251(1) | 3.37 | 2.66 | 0.09543 |
| 4.1(1) | 3.739(5) | 5.77(7) | 69.96(3) | 0.251(2) | 3.57 | 2.76 | 0.1009 |
| 5.1(1) | 3.723(6) | 5.71(9) | 68.67(4) | 0.251(5) | 4.28 | 3.48 | 0.1091 |
| 5.6(1) | 3.719(4) | 5.70(4) | 68.33(2) | 0.251(7) | 4.33 | 3.51 | 0.1095 |
| 6.0(1) | 3.714(6) | 5.68(7) | 67.95(7) | 0.251(8) | 3.28 | 2.67 | 0.09661 |
| 6.5(1) | 3.707(9) | 5.66(3) | 67.43(1) | 0.251(8) | 2.26 | 1.78 | 0.0597 |
| 7.0(1) | 3.702(9) | 5.65(1) | 67.09(7) | 0.252(1) | 3.47 | 2.74 | 0.09212 |
| 9.0(1) | 3.680(6) | 5.57(4) | 65.39(9) | 0.252(3) | 1.50 | 1.13 | 0.02936 |
| 9.4(1) | 3.676(2) | 5.56(2) | 65.09(3) | 0.252(4) | 0.97 | 0.74 | 0.0185 |
| 10.0(1) | 3.671(1) | 5.55(2) | 64.79(4) | 0.252(5) | 1.84 | 1.13 | 0.02604 |
| 10.5(1) | 3.667(6) | 5.53(1) | 64.43(6) | 0.252(0) | 0.67 | 0.5 | 0.01345 |
| 13.3(1) | 3.640(9) | 5.46(1) | 62.69(7) | 0.251(9) | 0.60 | 0.43 | 0.01293 |
| 14.3(1) | 3.630(3) | 5.43(9) | 62.08(8) | 0.251(9) | 1.68 | 1.31 | 0.03338 |
| 16.2(1) | 3.613(6) | 5.38(4) | 60.88(5) | 0.253(1) | 2.01 | 1.56 | 0.0481 |
| 18.4(1) | 3.600(2) | 5.35(4) | 60.10(1) | 0.253(4) | 1.53 | 1.22 | 0.03715 |
| 23.3(1) | 3.565(3) | 5.26(1) | 57.90(7) | 0.255(9) | 2.47 | 1.83 | 0.08125 |
| 25.8(1) | 3.542(1) | 5.23(1) | 56.84(3) | 0.255(9) | 2.74 | 2.08 | 0.09246 |
| Decompression (5.4 GPa) | 3.738(3) | 5.65(6) | 68.41(4) | 0.254(7) | 1.27 | 0.94 | 0.02951 |

Table 2: Structural data as a function of pressure for trigonal SnS$_2$ taken from Filso *et al.* [11]

Table 1 Selected crystal structure data at the 14 different pressure conditions explored. The refined space group was $P\bar{3}m1$ for the entire pressure range

| p (GPa) | a (Å) | c (Å) | V (Å$^3$) | z (S) | d(Sn–S) (Å) | d(S⋯S) (Å) | ∠(Sn–S–Sn) (°) |
|---|---|---|---|---|---|---|---|
| 0.0001 | 3.6456(4) | 5.8934(11) | 67.83(2) | 0.2473(3) | 2.5601(11) | 3.647(3) | 90.80(7) |
| 0.664(9) | 3.6390(3) | 5.7853(19) | 66.35(2) | 0.2513(15) | 2.555(5) | 3.563(10) | 90.82(16) |
| 1.109(11) | 3.6339(2) | 5.7322(17) | 65.56(2) | 0.2544(16) | 2.555(6) | 3.511(11) | 90.65(18) |
| 1.991(16) | 3.6215(3) | 5.650(3) | 64.18(3) | 0.2565(14) | 2.544(5) | 3.456(9) | 90.76(17) |
| 3.40(3) | 3.6067(3) | 5.544(2) | 62.46(3) | 0.2631(14) | 2.542(5) | 3.352(9) | 90.36(15) |
| 4.22(3) | 3.5983(4) | 5.494(3) | 61.61(3) | 0.2639(10) | 2.533(4) | 3.324(7) | 90.50(13) |
| 6.29(5) | 3.5766(4) | 5.393(3) | 59.74(3) | 0.2701(13) | 2.527(5) | 3.227(8) | 90.09(14) |
| 8.08(6) | 3.5605(4) | 5.317(3) | 58.37(3) | 0.2729(16) | 2.516(5) | 3.171(10) | 90.07(16) |
| 9.78(8) | 3.5455(3) | 5.257(3) | 57.23(3) | 0.2738(18) | 2.502(6) | 3.138(11) | 90.21(17) |
| 11.88(9) | 3.5388(3) | 5.157(2) | 56.13(3) | 0.2775(15) | 2.497(5) | 3.079(9) | 90.23(15) |
| 13.68(19) | 3.5362(4) | 5.100(3) | 55.23(3) | 0.2836(15) | 2.502(5) | 3.007(8) | 89.93(15) |
| 14.8(2) | 3.5343(4) | 5.068(3) | 54.82(3) | 0.2839(12) | 2.497(4) | 2.994(7) | 90.11(13) |
| 16.8(4) | 3.5272(4) | 5.026(3) | 54.16(3) | 0.2859(17) | 2.492(5) | 2.963(9) | 90.08(16) |
| 20(1) | 3.5074(7) | 4.978(4) | 53.05(5) | 0.2879(19) | 2.481(6) | 2.926(10) | 89.97(17) |

FIGURES

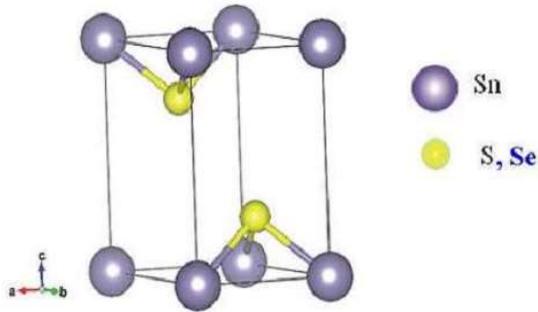

Figure 1: Conventional unit cell of trigonal $SnSe_2$ and $SnS_2$.

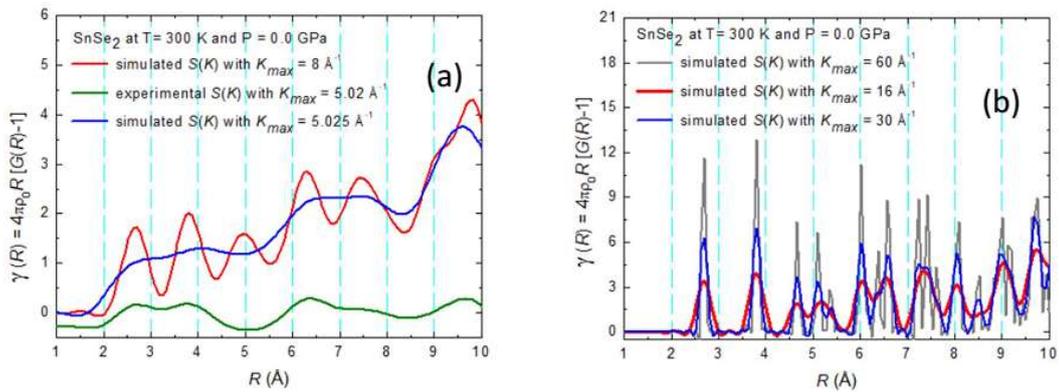

Figure 2: Experimental pair distribution function $G(R)$ of trigonal $SnSe_2$ at atmospheric pressure and simulated with different values of the transferred momentum $K$ vector.

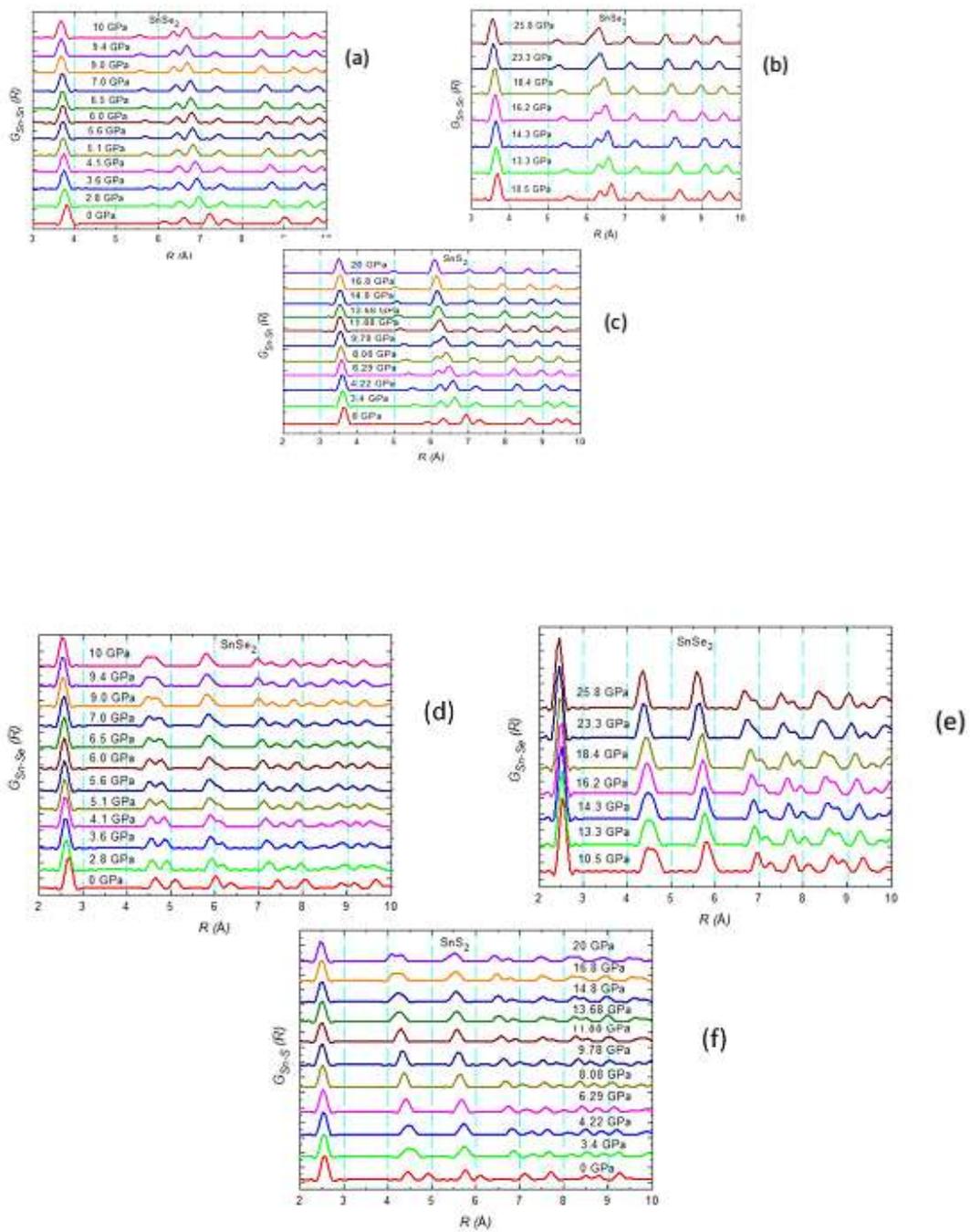

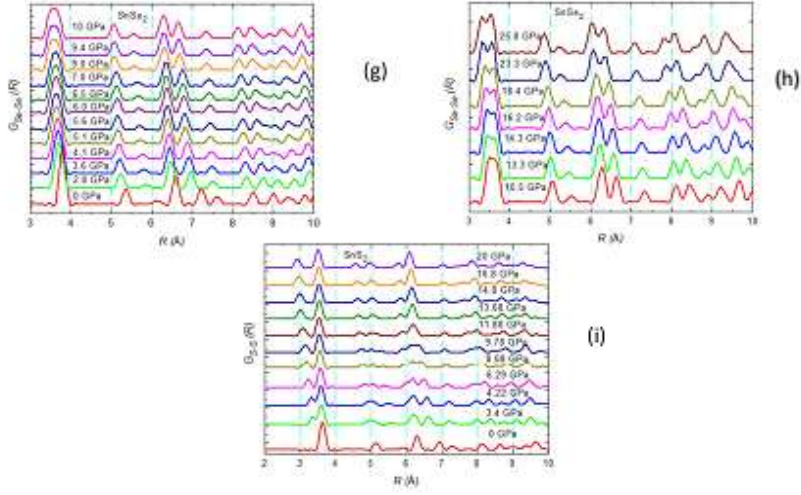

Figures 3(a)-(c), (d)-(f) and (g)-(i): Pressure dependence of the simulated $G_{Sn-Sn}(R)$, $G_{Sn-Se}(R)$, $G_{Sn-S}(R)$, $G_{Se-Se}(R)$, and $G_{S-S}(R)$ functions for trigonal $SnSe_2$ and $SnS_2$, with Kmax = 30 Å$^{-1}$.

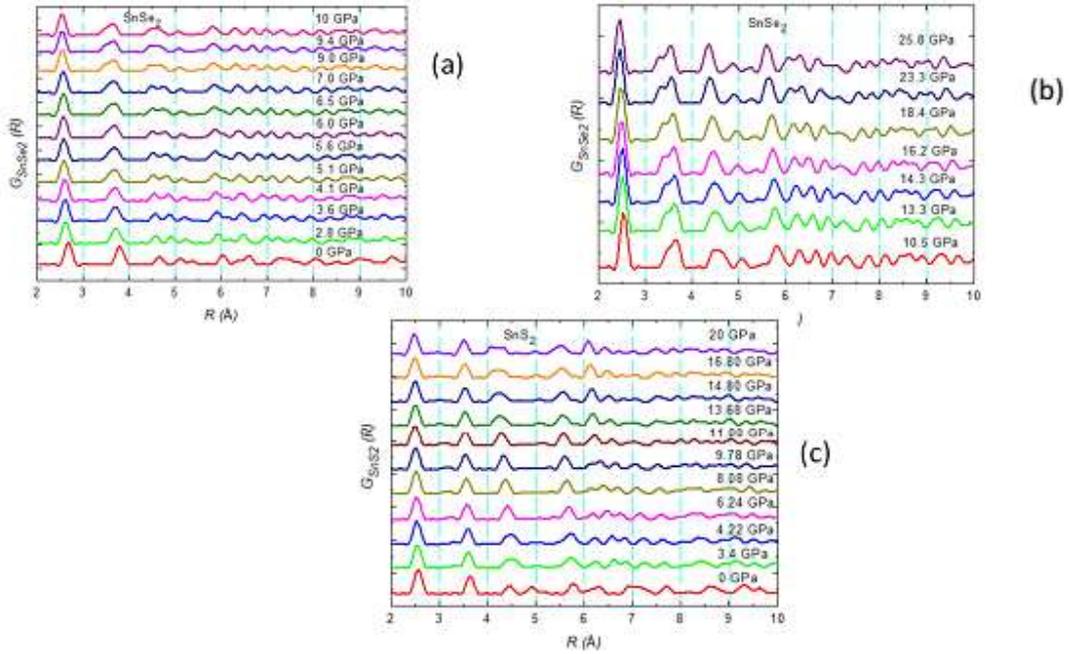

Figure 4(a)-(c): Pressure dependence of the simulated $G_{SnSe2}$ and $G_{SnS2}(R)$ functions for trigonal $SnSe_2$ and $SnS_2$, with Kmax = 30 Å$^{-1}$.

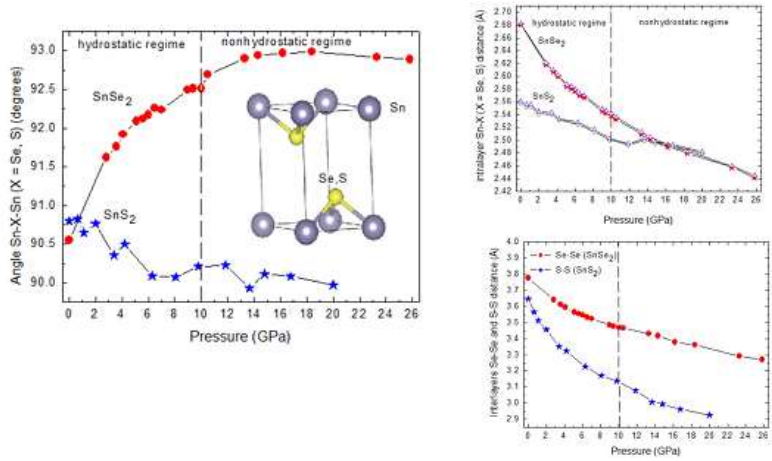

Figure 5: Pressure dependence of angle Sn-X (X=Se,S)-Sn and of intralayer distances Sn-X(X=Se,S) and interlayers distances X-X (X=Se,S).

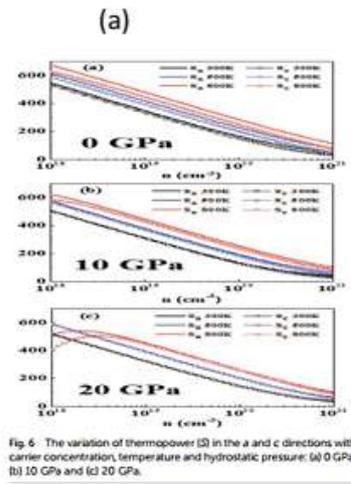
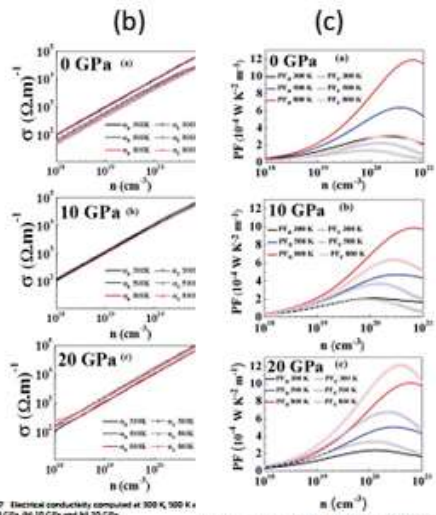
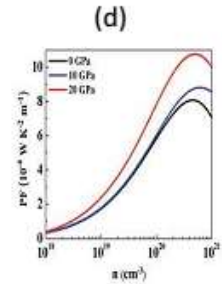
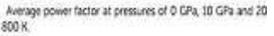

Figure 6(a)-(d): Pressure dependence of the thermoelectric power (S), electrical conductivity (σ), power fator (PF), and average power fator at atmospheric pressure, 10 GPa, and 20 GPa at temperatures 300 K, 500 K and 800 K, taken from Ref. [12].